\documentclass{ws-procs9x6}

\begin{document}

\def\ie{{\it i.e.}}
\def\cf{{\it cf.}}
\def\<{\langle}
\def\>{\rangle}
\def\len{{\mathcal L}}
\def\half{{1 \over 2}}
\def\quarter{{1 \over 4}}
\def\({\left(}
\def\){\right)}

\title{
\vspace{-5\baselineskip}
\begingroup
\footnotesize\normalfont\raggedleft
\lowercase{\sf hep-th/0401138} \\ 
SU-ITP-03/36\\
\vspace{\baselineskip}
\endgroup
Black hole singularity in AdS/CFT\footnote{
\uppercase{T}o appear in the \uppercase{P}roceedings of the 3rd 
\uppercase{S}ymposium on \uppercase{Q}uantum
\uppercase{T}heory and \uppercase{S}ymmetries (\uppercase{QTS}3),
\uppercase{C}incinnati, \uppercase{O}hio,
10--14 \uppercase{S}ept 2003 ---
\copyright\
\uppercase{W}orld \uppercase{S}cientific.}
}

\author{Veronika Hubeny}
\address{Department of Physics,
Stanford University,
Stanford, CA  94305,  USA \\
E-mail:  veronika@itp.stanford.edu}  

\maketitle

\abstracts{
We present a short review of hep-th/0306170.
In the context of AdS/CFT correspondence,
we explore what information from
behind the horizon of the bulk black hole geometry
can be found in boundary CFT correlators.
In particular, we argue that the CFT correlators 
contain distinct, albeit subtle, signals of the black
hole singularity.
}


It is a well-founded expectation that black holes should provide a key, or 
at least a window, into quantum gravity.  Typically, black holes contain 
curvature singularities, where classical laws of general relativity break 
down, to be replaced by  more fundamental quantum physics.  Unfortunately, 
an asymptotic observer cannot see this quantum singularity resolution in 
action, because the near-singularity physics is cloaked by an event 
horizon, the defining feature of a black hole.
The fact that black holes are causally nontrivial, signaled by an event 
horizon, has also associated with it long-standing puzzles, such as the 
information paradox.
 
To unravel the mysteries of black holes, we call to aid string theory, 
which has already proved very successful in resolving many
 timelike singularities.  
Spacelike singularities appear to be more subtle, and we expect that a fully 
nonperturbative formulation of string theory will be essential for this 
endeavor.  While there are several remarkable formulations, the one best 
suited for our purposes is the well-known AdS/CFT 
correspondence\cite{AdSCFT}.  
In particular, we will use the duality between 10-dimensional IIB string 
theory on $AdS_5 \times S^5$
and ${\mathcal N} = 4$, 4-dimensional Super Yang Mills gauge theory, 
living on the boundary of AdS.

More specifically, to study a singularity and an event horizon in the bulk 
gravity theory, we will consider the Schwarzschild-AdS geometry, 
describing a large black hole in AdS, whose gauge dual corresponds to an 
approximately thermal state.  By analyzing the properties of this state, 
such as the expectation values of various operators, 
we can hope to extract information about the black hole.  Indeed, due to 
the powerful nature of the AdS/CFT duality, we expect to be able to decode 
the full quantum near-singularity bulk physics from the gauge dual.

However, since such decoding requires that the gauge theory encodes 
behind-the-horizon physics in the first place,
we must first ask, {\it how much of the bulk physics does the gauge dual 
encode?}
Naive considerations might lead one
to think that the CFT encodes only the region of the bulk which is 
causally connected to the boundary, \ie\  outside the event horizon.
On the other hand, an event horizon is a global object (defined as the 
boundary of the past of the future infinity), which means that we cannot 
determine the presence or position of the event horizon without knowing 
the entire future evolution of the spacetime.
Hence if the horizon were to bound what the CFT can encode, then the 
AdS/CFT correspondence would have to be very nonlocal in time.

\begin{figure}[ht]
\centerline{\epsfxsize=3.in\epsfbox{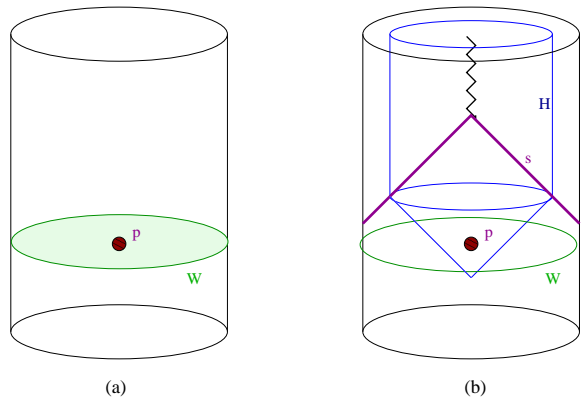}}   
\caption{Sketch of the proposed process of abstracting information from 
inside of the horizon: 
a) An event $p$, in locally pure AdS space is measured by $W$.
b) After this measurement is performed, a shell $s$ collapses, forming a 
black hole with a horizon $H$ which encompasses $p$. \label{inter}}
\end{figure}
We illustrate this point by the following gedanken-experiment\cite{prec} 
(\cf\ Fig.1):
Start with ``empty'' AdS space, and consider some event, labeled $p$ in 
Fig.1a.   As a boundary observer, one can obtain instantaneous information 
about the event $p$, for instance by measuring appropriately decorated 
Wilson loop\cite{wilsprec} $W$.
{\it After} this measurement has been made, i.e.\ entirely in the future 
of $W$, one can send in a shell $s$ of radiation with sufficient energy 
such that when it implodes in the center of AdS, it forms a large black 
hole, as sketched in Fig.1b.
The main observation is that
the global event horizon for this spacetime,  denoted by $H$ in Fig.1b, 
can originate at the center of AdS prior to $p$.
In particular, provided the shell is sent in within time of order the AdS 
radius after the measurement of $p$ is performed, one can always make the 
resulting black hole large enough for its horizon to encompass  $p$. 
This implies that one has in fact succeeded in ``measuring'' an 
event, $p$, which is inside a black hole horizon.

While the above argument indicates that the CFT should encode at least 
some physics inside a horizon, it does not lend itself to detailed 
computational analysis.  Instead, it is more fruitful to turn to 
a simpler set-up, namely that of the eternal black hole, which is static 
outside the horizon and has both future and past singularities, as well as
{\it two} boundaries.  This geometry was analyzed previously in 3 
dimensions by a number of authors\cite{Louko,Maldacena,KOS}; here we 
study the higher-dimensional analog\cite{sing}.

The real-time or thermofield formalism for thermal field 
theory\cite{realtime} is especially well-suited for this analysis.  One 
copy of the CFT resides on each of the two asymptotic regions; the CFTs 
are noninteracting but entangled through the Hartle-Hawking 
state\cite{Maldacena}.
In this approach, the one boundary thermal description is recovered by 
tracing over the Hilbert space of the other boundary CFT. 

Moreover, as demonstrated previously for 3 dimensions\cite{KOS}, 
one 
can  probe physics behind the horizon by studying 
the correlator of two operators, one on each asymptotic
boundary, each creating a large mass bulk particle.
As the mass $m \rightarrow \infty$, the correlator can be evaluated in the 
semi-classical geodesic approximation and is given by $\exp(-m \len)$, 
where $\len$ is the (regularized) proper length of the spacelike geodesic  
joining the boundary points. Because the geodesic passes through spacetime 
regions inside
 the horizon, this boundary correlator reveals information about the 
geometry behind the horizon.  

Unfortunately, the three dimensional case, which is simple enough to study 
analytically, is rather special.  The geometry is locally that of pure 
$AdS_3$, and the black hole singularity is merely a result of the orbifold 
nature of this geometry\cite{BTZ}.  Consequently, the geodesics 
are not sensitive to the position of the singularity, and correspondingly 
the correlation function is relatively structureless. 

On the other hand, this situation changes drastically in higher 
dimensions.  In particular, the singularity in  $d>3$ dimensions 
approaches that of the $d$-dimensional Schwarzschild black hole 
singularity, which has a dramatic effect on the spacelike geodesics:
pictorially, spacelike radial geodesics ``bounce off'' from the singularity.
For sake of simplicity, we focus on $d=5$ where the boundary CFT is four 
dimensional ${\mathcal N }=4$ SYM;
however, our results are qualitatively similar to those in 
other higher dimensions as well.

The first surprise, underlining the marked difference from the 
previously-studied 3-D case, 
is that the Penrose diagram\footnote{ 
Penrose diagram is a conformal diagram of a spacetime including its boundary,
used to describe its causal properties; time runs
up and space runs sideways, such that null rays lie at 45 degree angles.}
of the Schw-AdS$_5$ spacetime is different.
While the Penrose diagram for the Schw-AdS$_3$ geometry has the shape of 
a square, this can no longer be true in higher dimensions\cite{sing}.
One reason, apparent from Fig.2,
 is that outgoing radial null geodesics starting at the past
singularity do not reach the boundary at a time-symmetric point.
\begin{figure}[ht]
\centerline{\epsfxsize=3.9in\epsfbox{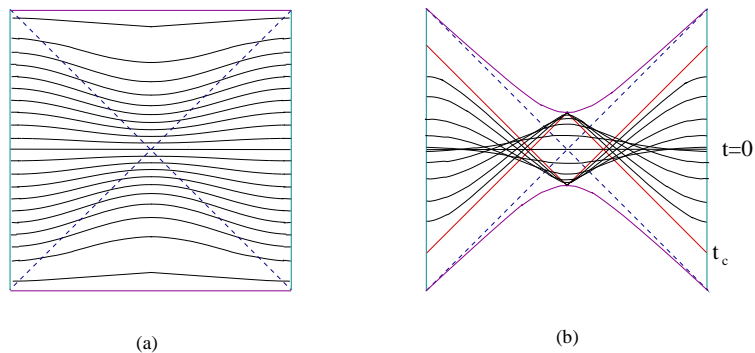}}   
\caption{Penrose diagrams of Schw-AdS in (a) 3 and (b) 5 dimensions.
The top and bottom curves are the future and past singularities, the vertical 
lines are the boundaries, the dashed diagonal lines are the
 event horizons, and 
the remaining curves are spacelike radial geodesics.
\label{inter}}
\end{figure}
Apart from the differences in the Penrose diagrams, there are two important
differences in the symmetric 
radial spacelike geodesics:  
Unlike the 3-D case, in higher dimensions there exists 
a particular time $t_c$ beyond which there are no geodesics connecting 
the two boundaries, and the geodesics cross each other.


This has an important
consequence for our present goal of seeing a signature of the singularity
in the dual field theory.  
Consider the boundary to boundary correlator  $\< \phi \phi \> (t)$
between two high-dimension 
operators inserted  in a symmetric fashion on the two boundaries
at time $t$.
Since this correlation function
 is related to the  proper length of a spacelike
geodesic connecting the two points, 
we expect that it reflects the special
behaviour of the geodesics as $ t \to t_c $.
In fact, direct evaluation suggests that we should see a pole in the 
correlation function:
\begin{equation}
\< \phi \phi \> (t) \sim e^{-m \len(t)} \sim {1 \over (t-t_c)^{2m}}
\qquad {\rm as} \qquad  t \to t_c \ .
\label{eq:correl}
\end{equation}
This corresponds to a light-cone singularity in the field theory,
since the geodesics are becoming almost null.

Had this been the full story, we would have a glaring signature of the
black hole singularity in the gauge theory dual.
However, general considerations of the
boundary field theory rule this out; one can easily show
that  $| \< \phi \phi \> (t) | \le | \< \phi \phi \> (0) | < \infty$. 
What went wrong?  In evaluating Eq.(\ref{eq:correl})
we assumed that the correlator is dominated by a single geodesic, 
namely the real, ``bounce'' geodesic shown in Fig.2b.
But, in fact, this
geodesic does {\it not} dominate the correlator, for there are in
general multiple geodesics that connect the two boundary points.
One indicator of this fact is the intersection of nearby geodesics
around $t=0$; in a cut-off field theory there would therefore be 
{\it three} geodesics connecting the two points at (cut-off) boundaries.  
Alternately, in a Euclidean picture, 
one real and two purely imaginary geodesics contribute.
At $t=0$ their proper distances coincide, creating a 
branch point\footnote{
The full analytic structure of the correlator is given by the 3-sheeted
Riemann surface of $\len(t)$, which can be derived implicitly
in terms of ``energy'' $E$ 
of the real geodesics;
$$
t = \half \ln \({ \half E^2 - E + 1 \over \sqrt{1 + \quarter E^4}}\)
- \half \, i \, 
       \ln \({- \half E^2 + i E + 1 \over \sqrt{1 + \quarter E^4}}\) 
\  , \ \ \ \
\len = \ln \({2 \over \sqrt{1 + \quarter E^4}}\)
$$}
in the correlator which behaves as $t^{4/3}$ for small $t$. 
(The 3 in the denominator of the exponent, implying a 3-sheeted
Riemann surface, corresponds to the 3 geodesics.)

By studying various resolutions of this branch point, we show that as
$t$ increases from $0$, the correlator defined by the boundary CFT
is given by a symmetric sum of the two complex branches of this
expression. Each of these can be attributed to a complex geodesic
in complexified spacetime. 
But the correlator is an analytic function of $t$ and can be
continued onto the real sheet. 
This is the essential feature of the CFT which enables us to extract
information which is not directly measurable.
On the real sheet, the ``light cone singularity" does appear. So the
boundary correlator {\it does} contain information about the
singularity, albeit in a subtle way.

So far, our discussion dealt with infinitely massive field $\phi$, 
as well as classical spacetime with vanishing string length $l_s$
and coupling $g_s$.  We can extract further information about
the singularity by deviating away from this limit.
Although the branch point at $t=0$ is smoothed out for any
finite $m$, preventing analytic continuation to the bouncing
geodesic, at each order in the $1/m$ expansion the branch
point persists and one can follow this geodesic and the
accompanying fluctuation corrections into the bouncing domain. 
Moreover, since the $1/m$ corrections are largest near the singularity,
we can effectively isolate the near-singularity behaviour.

One may similarly  make use of the  $l_s$ and $g_s$ corrections,
thereby extracting stringy and quantum near-singularity behaviour.
The main corrections at finite but small $l_s$ (with $g_s$ kept
equal to zero)  can be expressed as modifications of the
supergravity field equations, which does not change the
basic picture outlined above:
There still exists a branch point, allowing analytic continuation 
onto the real sheet.
The $g_s$ expansion is more analogous to the $1/m$ expansion;
finite $g_s$ smooths out the singularity.  But by
taking appropriate $g_s \to 0 $ limits we can extract
behavior around the singularity to all orders in $g_s$, as well as
study certain leading nonperturbative effects.


In summary, we have seen that the CFT encodes physics behind the
horizon, including the near-singularity region.
Unlike the previously-studied three dimensional case, in higher 
dimensions there is a genuine curvature singularity which is
bowed in on the Penrose diagram, leading to a special behaviour 
of the geodesics as a critical time $t_c$ is approached. 
Despite subtleties related to the nontrivial analytic structure of 
$\len(t)$, the CFT correlators reveal distinct signals of the
black hole singularity.  In fact, given\footnote{
Of course, it is not presently feasible to obtain the CFT ``data''
 $\< \phi \phi \>$ in the relevant regime by a direct computation:
the CFT is strongly coupled, and $m \to \infty$ implies infinitely
large dimension operators.  Nevertheless, it is still of use to ask
the matter-of-principle question:  with  given information 
about the CFT, what can one learn about the bulk?
}
the CFT data $\< \phi \phi \>$, the properties of the singularity
are computationally accessible\cite{sing}.
Since the ``$t_c$'' 
singularity persists to all orders in ${1 \over m}$ and $g_s$,
as well as for small $l_s$, we can also extract the stringy and quantum
behaviour near the black hole singularity.
Thus, using analyticity, we have demonstrated 
that a significant amount of information
from behind the horizon, and in particular from near the
singularity, is encoded in  boundary theory correlators.
Since analyticity was essential for our arguments, it would be 
worthwhile to understand its implications in the context of AdS/CFT
at a much deeper level.


\section*{Acknowledgments}

I wish to thank my collaborators, Lukasz Fidkowski, Matthew Kleban, and 
Stephen Shenker.  This work is supported in part by  NSF grant
PHY-9870115 and by the Stanford Institute for Theoretical Physics.


\noindent
\small
Due to space restrictions, only a very limited set of references is listed
here.  Please refer to Ref.[\refcite{sing}] for a more complete list.


\begin{thebibliography}{0}

\bibitem{AdSCFT} J.~M.~Maldacena,
{\it Adv.\ Theor.\ Math.\ Phys.}  {\bf 2}, 231 (1998)
[arXiv:hep-th/9711200].
E.~Witten,
{\it Adv.\ Theor.\ Math.\ Phys.}  {\bf 2}, 253 (1998)
[arXiv:hep-th/9802150].
O.~Aharony, S.~S.~Gubser, J.~M.~Maldacena, H.~Ooguri and Y.~Oz,
{\it Phys.\ Rept.}  {\bf 323}, 183 (2000)
[arXiv:hep-th/9905111].

\bibitem{prec} V.~E.~Hubeny,
{\it Int.\ J.\ Mod.\ Phys.\ D} {\bf 12}, 1693 (2003)
[arXiv:hep-th/0208047].

\bibitem{wilsprec} L.~Susskind and N.~Toumbas,
{\it Phys.\ Rev.\ D} {\bf 61}, 044001 (2000)
[arXiv:hep-th/9909013].
B.~Freivogel, S.~B.~Giddings and M.~Lippert,
{\it Phys.\ Rev.\ D} {\bf 66}, 106002 (2002)
[arXiv:hep-th/0207083].

\bibitem{Louko} V.~Balasubramanian and S.~F.~Ross,
{\it Phys.\ Rev.\ D} {\bf 61}, 044007 (2000)
[arXiv:hep-th/9906226].
J.~Louko, D.~Marolf and S.~F.~Ross,
{\it Phys.\ Rev.\ D} {\bf 62}, 044041 (2000)
[arXiv:hep-th/0002111].

\bibitem{Maldacena} J.~M.~Maldacena,
{\it JHEP} {\bf 0304}, 021 (2003)
[arXiv:hep-th/0106112].

\bibitem{KOS} P.~Kraus, H.~Ooguri and S.~Shenker,
{\it Phys.\ Rev.\ D} {\bf 67}, 124022 (2003)
[arXiv:hep-th/0212277].

\bibitem{sing} L.~Fidkowski, V.~Hubeny, M.~Kleban and S.~Shenker,
arXiv:hep-th/0306170.

\bibitem{realtime} M.~Le Bellac, {\it Thermal Field Theory}, Cambridge 
University Press (1966).

\bibitem{BTZ} M.~Banados, C.~Teitelboim and J.~Zanelli,
{\it Phys.\ Rev.\ Lett.}  {\bf 69}, 1849 (1992)
[arXiv:hep-th/9204099].

\end{thebibliography}
\end{document}